\documentclass[preprint,amsmath,amssymb,aps,showkeys,showpacs]{revtex4}
\usepackage[english]{babel}
\usepackage{graphicx}
\usepackage{graphics}
\usepackage{amsmath}
\usepackage{dcolumn}
\usepackage{amssymb}
\usepackage{bm}
\usepackage[latin1]{inputenc}

\begin{document}
\title{On the influence of a Coulomb-like potential induced by the Lorentz symmetry breaking effects on the Harmonic Oscillator}
\author{K. Bakke}
\email{kbakke@fisica.ufpb.br}
\affiliation{Departamento de F\'isica, Universidade Federal da Para\'iba, Caixa Postal 5008, 58051-970, Jo\~ao Pessoa, PB, Brazil.}

\author{H. Belich} 
\affiliation{Departamento de F\'isica e Qu\'imica, Universidade Federal do Esp\'irito Santo, Av. Fernando Ferrari, 514, Goiabeiras, 29060-900, Vit\'oria, ES, Brazil.}

\begin{abstract}
In this work, we obtain bound states for a nonrelativistic spin-half neutral particle under the influence of a Coulomb-like potential induced by the Lorentz symmetry breaking effects. We present a new possible scenario of studying the Lorentz symmetry breaking effects on a nonrelativistic quantum system defined by a fixed space-like vector field parallel to the radial direction interacting with a uniform magnetic field along the $z$-axis. Furthermore, we also discuss the influence of a Coulomb-like potential induced by Lorentz symmetry violation effects on the two-dimensional harmonic oscillator. 
\end{abstract}

\maketitle

\section{Introduction}

Symmetries are fundamental guides when one intends to systematize the study of any theory. In this sense, Lorentz and CPT invariances acquire supreme importance in modern Quantum Field Theory, both symmetries being respected by the Standard Model for Particle Physics. Discussions about symmetry breaking are well-known in nonrelativistic quantum systems involving phase transitions such as ferromagnetic systems, where the rotation symmetry is broken due to the influence of a magnetic field. For relativistic systems, the study of symmetry breaking can be extended by considering a background given by a $4$-vector field that breaks the symmetry $\mathcal{SO}\left( 1,3\right) $ instead of the symmetry $\mathcal{SO}\left( 3\right)$. This new possibility of spontaneous violation was first suggested by Kostelecky and Samuel \cite{extra3} in 1989 indicating that, in the string field theory, the spontaneous violation of symmetry by a scalar field could be extended. This line of research is known in the literature as the spontaneous violation of the Lorentz symmetry \cite{extra3,extra1,extra2}. In the electroweak theory, a scalar field acquires a nonzero vacuum expectation value which yields mass to gauge bosons (Higgs Mechanism). Similarly, in the string field theory, this scalar field can be extended to a ``tensor'' field. Nowadays, these theories are encompassed in the framework of the Standard Model Extension (SME) as a possible extension of the minimal Standard Model of the fundamental interactions.

The gauge sector of SME \cite{extra3,Colladay} covers a CPT-odd and a CPT-even subsectors \cite{Jackiw,KM3,Kostelec}, which have been examined in several aspects concerning supersymmetry \cite{Susy}, vacuum Cherenkov radiation emission \cite{Cherenkov2}, radiative corrections \cite{Radiative}, Casimir effect \cite{Casimir}, anisotropies of the Cosmic Microwave Background Radiation \cite{CMBR}, and between other interesting issues \cite{Petrov}. Further, several works were devoted to examine consistency aspects of the CPT-odd subsector \cite{Adam,Soldati} and of the CPT-even one \cite{Prop2,Klinkmicro}. The CPT-even gauge sector of SME is obtained by including the gauge sector in the Lagrangian term, $\left(-\frac{1}{4}\left( K_{F}\right)_{\mu\nu\lambda\kappa}F^{\mu\nu}F^{\lambda\kappa}\right)$, with $\left(K_{F}\right)_{\mu\nu\lambda\kappa}$ being a Lorentz-violating tensor. The tensor $\left(K_{F}\right)_{\mu\nu\lambda\kappa}$ is composed of 19 coefficients, where nine of these coefficients are nonbirefringent and ten are birefringent, all of them endowed with the symmetries of the Riemann tensor and a double null trace $\left(K_{F}\right)^{\mu\nu}_{\,\,\,\,\,\,\mu\nu}=0$. The effects of this CPT-even electrodynamics on the fermion-fermion interaction was considered in Refs. \cite{Klink3,Interact2,Interact3}.

The Standard Model is believed to be a low-energy effective theory from a underlying unified description of gravity and quantum physics at the Planck scale. Taking into account that a spontaneous Lorentz violations occurs beyond the electroweak scale, the search for a theory of quantum gravity necessarily must include such violation. With this motivation, one can analyse neutral fermions in a general coordinate system in order to investigate the influence of the Lorentz violation on new properties of neutrinos moving in a curved spacetime. An appropriate treatment to study possible extensions of the Standard Model to a curved spacetime is via the tetrads formalism. SME was presented in the context of a Riemann-Cartan spacetime in Ref. \cite{tet}. In the matter sector of SME, Dirac spinor fields can be used for describing the matter-gravity couplings that are expected to dominate in many experimental scenarios. A new possibility of experimental verification of the Lorentz violation in these scenarios is promoted by background fields \cite{bbs2,bb}.

The modified Dirac theory has already been examined in the literature \cite{Hamilton}, and, in the nonrelativistic limit of the modified Dirac theory, the spectrum of energy of the hydrogen atom has been discussed in \cite{Manojr, Nonmini, novo, novo1}. Hence, there is an extensive amount of work looking at the Lorentz violation, and numerous experimental bounds exist \cite{extra2}. Recently, by introducing nonminimal couplings with the background of the Lorentz symmetry violation, geometric quantum phases \cite{ab,ahan,anan2,berry} have been obtained \cite{belich,belich1,belich2,belich3}. Moreover, the Aharonov-Casher effect \cite{ac} and the relativistic Anandan quantum phase \cite{anan2} have been studied in the context of the Lorentz symmetry breaking in \cite{bbs2}. This work has discussed that a new phase could detect the presence of a background field through interference experiments. Then, we could verify if the Lorentz violation describes neutrino oscillations, which makes of these measurements sensitive probes of new physics \cite{proc}. Other studies of the influence of a symmetry breaking background have been made for bound states for a relativistic neutral particle with an analogue of the permanent magnetic dipole moment in the Lorentz symmetry violation background \cite{bbs}, and for a charged particle describing a circular path in presence of a Lorentz-violating background nonminimally coupled to a spinor and a gauge field \cite{mano}.

In this work, we obtain bound states for a nonrelativistic spin-half neutral particle under the influence of a Coulomb-like potential induced by the Lorentz symmetry breaking effects. Here, we present a new possible scenario of studying Lorentz symmetry breaking effects based on the assumption of a fixed space-like vector field background parallel to the radial direction. Further, we also discuss the influence of a Coulomb-like potential induced by Lorentz symmetry violation effects on the two-dimensional harmonic oscillator.

This paper is organized as follows: in section II, we present the Lorentz symmetry violation background and give a brief review of the nonrelativistic quantum dynamics of a Dirac neutral particle in this scenario. In the following, we obtain the bound states solutions of the the Schr\"odinger-Pauli equation corresponding to having a  spin-half neutral particle under the influence of a Coulomb-like potential; in section III, we discuss the influence of the Coulomb-like potential induced by Lorentz symmetry breaking effects on the the two-dimensional harmonic oscillator; in section IV, we present our conclusions.

\section{Coulomb-like potential induced by Lorentz symmetry breaking effects}

In this section, we obtain the energy levels for bound states for a spin-half neutral particle under the influence of a Coulomb-like potential induced by the Lorentz symmetry breaking effects. Recently, we have proposed the study of the He-McKellar-Wikens effect \cite{hmw} and the scalar Aharonov-Bohm effect \cite{anan2} based on Lorentz symmetry breaking effects by modifying the nonminimal coupling proposed in \cite{belich1,belich} and writing it in the form:
\begin{eqnarray}
i\gamma^{\mu}\partial_{\mu}\rightarrow i\gamma^{\mu}\partial_{\mu}-g\,b^{\mu}\,F_{\mu\nu}\left(x\right)\,\gamma^{\nu},
\label{1}
\end{eqnarray}
where $g$ is a constant, and $b^{\mu}$ corresponds to a fixed $4$-vector that acts as a vector field breaking the Lorentz symmetry violation. The tensor $F_{\mu\nu}\left(x\right)$ corresponds to the usual electromagnetic tensor ($F_{0i}=-F_{i0}=-E_{i}$, and $F_{ij}=-F_{ji}=\epsilon_{ijk}B^{k}$), and the $\gamma^{\mu}$ matrices are defined in the Minkowski spacetime in the form \cite{greiner}:
\begin{eqnarray}
\gamma^{0}=\hat{\beta}=\left(
\begin{array}{cc}
1 & 0 \\
0 & -1 \\
\end{array}\right);\,\,\,\,\,\,
\gamma^{i}=\hat{\beta}\,\hat{\alpha}^{i}=\left(
\begin{array}{cc}
 0 & \sigma^{i} \\
-\sigma^{i} & 0 \\
\end{array}\right);\,\,\,\,\,\,\Sigma^{i}=\left(
\begin{array}{cc}
\sigma^{i} & 0 \\
0 & \sigma^{i} \\	
\end{array}\right),
\label{2}
\end{eqnarray}
with $\vec{\Sigma}$ being the spin vector. The matrices $\sigma^{i}$ correspond to the Pauli matrices, and satisfy the relation $\left(\sigma^{i}\,\sigma^{j}+\sigma^{j}\,\sigma^{i}\right)=2\eta^{ij}$. In this way, the Dirac equation (in Cartesian coordinates) in the Lorentz symmetry violation background described by the nonminimal coupling (\ref{1}) is given by \cite{bbs3}
\begin{eqnarray}
i\frac{\partial\Psi}{\partial t}=m\hat{\beta}\Psi+\vec{\alpha}\cdot\vec{p}\,\psi-g\,b^{0}\vec{\alpha}\cdot\vec{E}\Psi-g\,\vec{\alpha}\cdot\left(\vec{b}\times\vec{B}\right)\Psi+g\,\vec{b}\cdot\vec{E}\Psi.
\label{3}
\end{eqnarray}

Our interest in this section is to discuss the nonrelativistic quantum dynamics of a spin-half neutral particle under the influence of a Coulomb-like potential induced by Lorentz symmetry breaking effects. As discussed in Refs. \cite{schu,bbs3,bbs}, if we intend to work with curvilinear coordinates, then, we need to apply a coordinate transformation $\frac{\partial}{\partial x^{\mu}}=\frac{\partial \bar{x}^{\nu}}{\partial x^{\mu}}\,\frac{\partial}{\partial\bar{x}^{\nu}}$, and a unitary transformation on the wave function $\psi\left(x\right)=U\,\psi'\left(\bar{x}\right)$. In this way, The Dirac equation (\ref{3}) can be written in any orthogonal system in the presence of Lorentz symmetry breaking effects described in (\ref{1}) as  
\begin{eqnarray}
i\,\gamma^{\mu}\,D_{\mu}\,\Psi+\frac{i}{2}\,\sum_{k=1}^{3}\,\gamma^{k}\,\left[D_{k}\,\ln\left(\frac{h_{1}\,h_{2}\,h_{3}}{h_{k}}\right)\right]\Psi-g\,b^{\mu}\,F_{\mu\nu}\left(x\right)\,\gamma^{\nu}=m\Psi,
\label{4}
\end{eqnarray}
where $D_{\mu}=\frac{1}{h_{\mu}}\,\partial_{\mu}$ is the derivative of the corresponding coordinate system, and the parameter $h_{k}$ correspond to the scale factors of this coordinate system \cite{schu}. For instance, the line element of the Minkowski spacetime is writing in cylindrical coordinates in the form: $ds^{2}=-dt^{2}+d\rho^{2}+\rho^{2}d\varphi^{2}+dz^{2}$; then, the corresponding scale factors are $h_{0}=1$, $h_{1}=1$, $h_{2}=\rho$ and $h_{3}=1$. Moreover, the second term in (\ref{4}) gives rise to a term called the spinorial connection \cite{schu,b4,bbs,bbs2,bbs3,bd,weinberg}. Hence, after simple calculations, we obtain the nonrelativistic limit of the Dirac equation (\ref{4}) in cylindrical coordinates, which is given by the following Schr\"odinger-Pauli equation \cite{bbs3}
\begin{eqnarray}
i\frac{\partial\psi}{\partial t}=\frac{1}{2m}\left[\vec{p}-i\vec{\xi}-g\,b^{0}\,\vec{E}-g\,\left(\vec{b}\times\vec{B}\right)\right]^{2}\psi+g\,\vec{b}\cdot\vec{E}\,\psi-\frac{g}{2m}\,\vec{\sigma}\cdot\vec{B}_{\mathrm{eff}}\,\psi,
\label{5}
\end{eqnarray} 
where the vector $\vec{\xi}$ comes from the contribution of the spinorial connection, and it is defined in such a way that its components are given by: $-i\xi_{k}=-\frac{\sigma^{3}}{2\rho}\,\delta_{2k}$. Moreover, we also have the presence of an effective magnetic field $\vec{B}_{\mathrm{eff}}$ which is defined as \cite{bbs3}
\begin{eqnarray}
\vec{B}_{\mathrm{eff}}=\vec{\nabla}\times\left[b^{0}\,\vec{E}+\left(\vec{b}\times\vec{B}\right)\right].
\label{6}
\end{eqnarray}

From the Schr\"odinger-Pauli equation (\ref{5}), we have studied recently the arising of bound states for a nonrelativistic spin-half neutral particle under the influence of a Coulomb-like potential produced by the presence of a radial electric field and a fixed space-like vector field $\vec{b}$ parallel to the radial direction \cite{bbs3}. In this paper, we consider a fixed space-like vector field and a magnetic field defined as
\begin{eqnarray}
\vec{b}=b\,\hat{\rho};\,\,\,\,\,\vec{B}=B_{0}\,\hat{z},
\label{2.1}
\end{eqnarray}
where $b$ and $B_{0}$ are constants, and $\hat{\rho}$ and $\hat{z}$ corresponding to unit vectors in the radial and $z$ directions, respectively. One should note that the presence of this uniform magnetic field and the choice of a fixed space-like vector field being parallel to the radial direction gives rise to a new possible scenario for the measurement of the Lorentz symmetry breaking. In a recent work \cite{bbs3}, a Coulomb-like potential is yielded by the term $g\,\vec{b}\cdot\vec{E}$ in Eq. (\ref{5}), which plays the role of a scalar potential. In the present case, a Coulomb-like potential is induced by Lorentz symmetry effects through the term $g\,\left(\vec{b}\times\vec{B}\right)$ in Eq. (\ref{5}), which acts on the neutral particle as a vector potential. In the following, we show that the background defined in (\ref{2.1}) yields a Coulomb-like potential allowing us to obtain bound states solutions for the Schr\"odinger-Pauli equation (\ref{5}). In this way, by using (\ref{2.1}), the Schr\"odinger-Pauli equation in cylindrical coordinates becomes
\begin{eqnarray}
i\frac{\partial\psi}{\partial t}&=&-\frac{1}{2m}\left[\frac{\partial^{2}}{\partial\rho^{2}}+\frac{1}{\rho}\frac{\partial}{\partial\rho}+\frac{1}{\rho^{2}}\,\frac{\partial^{2}}{\partial\varphi^{2}}+\frac{\partial^{2}}{\partial z^{2}}\right]\psi+\frac{1}{2m}\frac{i\sigma^{3}}{\rho^{2}}\,\frac{\partial\psi}{\partial\varphi}\nonumber\\
[-2mm]\label{2.5}\\[-2mm]
&+&\frac{1}{8m\rho^{2}}\,\psi-i\frac{gbB_{0}}{m\rho}\frac{\partial\psi}{\partial\varphi}+\frac{\left(gbB_{0}\right)^{2}}{2m}\psi.\nonumber
\end{eqnarray}

We can see in (\ref{2.5}) that $\psi$ is an eigenfunction of $\sigma^{3}$, whose eigenvalues are $s=\pm1$. Thus, we can write $\sigma^{3}\psi_{s}=\pm\psi_{s}=s\psi_{s}$. Note that the operators $\hat{p}_{z}=-i\partial_{z}$ and $\hat{J}_{z}=-i\partial_{\varphi}$ \cite{schu} commute with the Hamiltonian of the right-hand side of (\ref{2.5}), therefore the solution of (\ref{2.5}) can be written in terms of the eigenvalues of the operator $\hat{p}_{z}$, and the $z$-component of the total angular momentum $\hat{J}_{z}$ \footnote{The discussion about the expression of the $z$-component of the total angular momentum operator in cylindrical coordinates was done in Ref. \cite{schu}. There, it has been shown that the $z$-component of the total angular momentum in cylindrical coordinates is given by $\hat{J}_{z}=-i\partial_{\varphi}$, where the eigenvalues are $\mu=l\pm\frac{1}{2}$.}: 
\begin{eqnarray}
\psi_{s}\left(t,\rho,\varphi,z\right)=e^{-i\mathcal{E}t}\,e^{i\left(l+\frac{1}{2}\right)\varphi}\,e^{ikz}\,G_{s}\left(\rho\right),
\label{2.6}
\end{eqnarray}
where $l=0,\pm1,\pm2,\ldots$, $k$ is a constant, and $G\left(\rho\right)$ is a function of the radial coordinate. Since Eq. (\ref{2.5}) is diagonal, then, substituting (\ref{2.6}) into the Schr\"odinger-Pauli equation (\ref{2.5}) we shall obtain two non-coupled radial equations for both spins $s=\pm1$. Therefore, in order to write these two non-coupled equation in a compact form, we label $G_{s}\left(\rho\right)$ and write:
\begin{eqnarray}
G_{s}''+\frac{1}{\rho}G_{s}'-\frac{\nu_{s}^{2}}{\rho^{2}}G_{s}-\frac{\delta}{\rho}\,G_{s}+\zeta^{2}\,G_{s}=0,
\label{2.6a}
\end{eqnarray}
where we have defined the parameters:
\begin{eqnarray}
\nu_{s}&=&l+\frac{1}{2}\left(1-s\right);\nonumber\\
\delta&=&2gbB_{0}\nu_{s}+sgbB_{0};\label{2.7}\\
\zeta^{2}&=&2m\mathcal{E}-k^{2}-\left(gbB_{0}\right)^{2}.\nonumber
\end{eqnarray}

Note that the fourth term of the left-hand side of Eq. (\ref{2.6a}) arises from the effects of the Lorentz symmetry breaking, and plays the role of a Coulomb-like potential. Now, let us discuss the asymptotic behavior of the radial equation (\ref{2.6a}). For $\rho\rightarrow\infty$, we have 
\begin{eqnarray}
G_{s}''+\zeta^{2}\,G_{s}=0;
\label{2.8}
\end{eqnarray}
thus, we can find either scattering states $\left(G_{s}\cong e^{i\zeta\rho}\right)$, or bound states $\left(G_{s}\cong e^{-\tau\rho}\right)$ if we consider $\zeta^{2}=-\tau^{2}$ \cite{mello}. Our interest in this work is to obtain bound states, then, the term proportional to $\delta$ behaves like an attractive potential by taking negative values of $\delta$, that is, by considering $\delta=-\left|\delta\right|$ \cite{mello}. This is possible for $g<0$ and $B_{0}>0$ (we consider $b$ being always a positive number) or $g>0$ and $B_{0}<0$. Therefore, both these choices yield an analogue of an attractive Coulomb potential. Thereby, we rewrite Eq. (\ref{2.6a}) in the form:
\begin{eqnarray}
G_{s}''+\frac{1}{\rho}\,G_{s}'-\frac{\nu_{s}^{2}}{\rho^{2}}G_{s}+\frac{\left|\delta\right|}{\rho}\,G_{s}-\tau^{2}\,G_{s}=0.
\label{2.9}
\end{eqnarray}

Next, let us make a change of variables given by $r=2\tau\rho$. Thus, the radial equation (\ref{2.9}) becomes
\begin{eqnarray}
G_{s}''+\frac{1}{r}\,G_{s}'-\frac{\nu_{s}^{2}}{r^{2}}\,G_{s}+\frac{\left|\delta\right|}{2\tau\, r}\,G_{s}-\frac{1}{4}\,G_{s}=0.
\label{2.10}
\end{eqnarray}
In order to obtain a regular solution at the origin for the second-order differential equation (\ref{2.10}), we can write
\begin{eqnarray}
G_{s}\left(r\right)=e^{-\frac{r}{2}}\,r^{\left|\nu_{s}\right|}\,M_{s}\left(r\right).
\label{2.11}
\end{eqnarray}
Substituting (\ref{2.11}) into (\ref{2.10}), we obtain the following second-order differential equation
\begin{eqnarray}
r\,M_{s}''+\left[2\left|\nu_{s}\right|+1-r\right]M_{s}'+\left[\frac{\left|\delta\right|}{2\tau}-\left|\nu_{s}\right|-\frac{1}{2}\right]M_{s}=0.
\label{2.12}
\end{eqnarray}

The second-order differential equation (\ref{2.12}) is known in the literature as the Kummer equation or the confluent hypergeometric equation \cite{abra}. The function $M_{s}\left(r\right)$ corresponds to the Kummer function of first kind which is defined as
\begin{eqnarray}
M_{s}\left(r\right)=M\left(\left|\nu_{s}\right|+\frac{1}{2}-\frac{\left|\delta\right|}{2\tau},2\left|\nu_{s}\right|+1,r\right).
\label{2.13}
\end{eqnarray} 
A finite radial wave function can be obtained when we impose the condition where the confluent hypergeometric series becomes a polynomial of degree $n$ (where $n=0,1,2,\ldots$). This occurs when $\left|\nu_{s}\right|+\frac{1}{2}-\frac{\left|\delta\right|}{2\tau}=-n$ \cite{abra,landau}. In this way, by taking $\left(-\tau^{2}\right)=2m\mathcal{E}-k^{2}-\left(gbB_{0}\right)^{2}$, we obtain
\begin{eqnarray}
\mathcal{E}_{n,\,l}=-\frac{1}{8m}\frac{\left[2gbB_{0}\nu_{s}+sgbB_{0}\right]^{2}}{\left[n+\left|\nu_{s}\right|+1/2\right]^{2}}+\frac{\left(gbB_{0}\right)^{2}}{2m}+\frac{k^{2}}{2m}.
\label{2.15}
\end{eqnarray}

The expression (\ref{2.15}) correspond to the nonrelativistic energy levels for a neutral particle under the influence of a Coulomb-like potential induced by the effects of the Lorentz symmetry violation background defined by the presence of a uniform magnetic field on the $z$ direction, and a fixed space-like vector field parallel to the radial direction.

\section{Influence of the Coulomb-like potential induced by Lorentz symmetry breaking effects on the Harmonic oscillator}

In this section, we discuss the influence of the Coulomb-like potential induced by Lorentz symmetry breaking effects obtained in the previous section on the two-dimensional harmonic oscillator $V\left(\rho\right)=\frac{1}{2}m\omega\rho^{2}$. By following the steps from Eq. (\ref{2.1}) to Eq. (\ref{2.7}), we have the following radial equation
\begin{eqnarray}
G_{s}''+\frac{1}{\rho}\,G_{s}'-\frac{\nu_{s}^{2}}{\rho^{2}}\,G_{s}-\frac{\delta}{\rho}\,G_{s}-m^{2}\omega^{2}\,\rho^{2}\,G_{s}+\zeta^{2}\,G_{s}=0,
\label{3.1}
\end{eqnarray}
where $\nu_{s}$, $\zeta$ and $\delta$ are defined in (\ref{2.7}). Here, we consider the general case involving $\delta$ and $\zeta$, in other words, we do not need to consider the values corresponding $\delta=-\left|\delta\right|$ and $\zeta^{2}=-\tau^{2}$ as in the previous section. Next, we consider a new change of variables given by: $\xi=\sqrt{m\omega}\,\rho$. Thus, we have
\begin{eqnarray}
G_{s}''+\frac{1}{\xi}\,G_{s}'-\frac{\nu_{s}^{2}}{\xi^{2}}\,G_{s}-\frac{\delta}{\sqrt{m\omega}\,\xi}\,G_{s}-\xi^{2}\,G_{s}+\frac{\zeta^{2}}{m\omega}\,G_{s}=0.
\label{3.2}
\end{eqnarray}

Therefore, in order that the radial wave function can be regular at the origin, the solution of the second order differential equation (\ref{3.2}) is given in the form:
\begin{eqnarray}
G_{s}\left(\xi\right)=e^{-\frac{\xi^{2}}{2}}\,\xi^{\left|\nu_{s}\right|}\,H_{s}\left(\xi\right).
\label{3.3}
\end{eqnarray}
Substituting (\ref{3.3}) into (\ref{3.2}), we obtain
\begin{eqnarray}
H_{s}''+\left[\frac{2\left|\nu_{s}\right|+1}{\xi}-2\xi\right]H_{s}'+\left[g-\frac{\delta}{\sqrt{m\omega}\,\xi}\right]H_{s}=0,
\label{3.4}
\end{eqnarray}
where $g=\frac{\zeta^{2}}{m\omega}-2-2\left|\nu_{s}\right|$. The function $H_{s}\left(\xi\right)$ which is solution of the second order differential equation (\ref{3.4}) is known as the Heun biconfluent function \cite{heun,eug,b6}:
\begin{eqnarray}
H_{s}\left(\xi\right)=H\left[2\left|\nu_{s}\right|,\,0,\,\frac{\zeta^{2}}{m\omega},\,\frac{2\delta}{\sqrt{m\omega}},\,\xi\right].
\label{3.5}
\end{eqnarray} 

Our goal is to obtain the energy levels for bound states for both signs of the Coulomb-like potential. We use the Frobenius method \cite{arf,f1} in order that the solution of Eq. (\ref{3.5}) can be written as a power series expansion around the origin:
\begin{eqnarray}
H_{s}\left(\xi\right)=\sum_{m=0}^{\infty}\,a_{m}\,\xi^{m}.
\label{3.11}
\end{eqnarray} 

Substituting the series (\ref{3.11}) into (\ref{3.5}), we obtain the recurrence relation:
\begin{eqnarray}
a_{m+2}=\frac{\delta}{\sqrt{m\omega}}\,\frac{a_{m+1}}{\left(m+2\right)\,\left(m+\bar{\alpha}+1\right)}-\frac{\left(g-2m\right)}{\left(m+2\right)\,\left(m+\bar{\alpha}+1\right)}\,a_{m},
\label{3.12}
\end{eqnarray}
where $\bar{\alpha}=2\left|\nu_{s}\right|+1$ and $a_{1}=\frac{\delta}{\bar{\alpha}\,\sqrt{m\omega}}\,a_{0}$. By starting with $a_{0}=1$ and using the relation (\ref{3.12}), we can calculate other coefficients of the power series expansion (\ref{3.11}). For instance,
\begin{eqnarray}
a_{2}&=&\left(\frac{\delta}{\sqrt{m\omega}}\right)^{2}\frac{1}{2\bar{\alpha}\left(\bar{\alpha}+1\right)}-\frac{g}{2\left(\bar{\alpha}+1\right)};\nonumber\\
[-2mm]\label{3.13}\\[-2mm]
a_{3}&=&\left(\frac{\delta}{\sqrt{m\omega}}\right)^{3}\,\frac{1}{6\bar{\alpha}\left(\bar{\alpha}+1\right)\left(\bar{\alpha}+2\right)}-\left(\frac{\delta}{\sqrt{m\omega}}\right)\frac{g}{6\bar{\alpha}\left(\bar{\alpha}+1\right)\left(\bar{\alpha}+2\right)}\nonumber\\
&-&\left(\frac{\delta}{\sqrt{m\omega}}\right)\,\frac{\left(g-2\right)}{3\bar{\alpha}\left(\bar{\alpha}+2\right)}.\nonumber
\end{eqnarray}

Note that the recurrence relation (\ref{3.12}) is valid for both signs of the Coulomb-like potential, that is, we can specify the signs of the Coulomb-like potential by making $\delta\rightarrow\pm\left|\delta\right|$ in (\ref{3.12}). Therefore, in order to obtain finite solutions everywhere, which represent bound state solutions, we need that the power series expansion (\ref{3.11}) or the Heun Biconfluent series becomes a polynomial of degree $n$. Through the expression (\ref{3.12}), we can see that the power series expansion (\ref{3.11}) becomes a polynomial of degree $n$ if we impose the conditions:
\begin{eqnarray}
g=2n\,\,\,\,\,\,\mathrm{and}\,\,\,\,\,\,a_{n+1}=0,
\label{3.13}
\end{eqnarray}
where $n=1,2,3,\ldots$, and $g=\frac{\zeta^{2}}{m\omega}-2\left|\nu_{s}\right|-2$. From the condition $g=2n$, we can obtain the expression for the energy levels for bound states:
\begin{eqnarray}
\mathcal{E}_{n,\,l,\,s}=\omega_{n,\,l,\,s}\left[n+\left|\nu_{s}\right|+1\right]+\frac{\left(gbB_{0}\right)^{2}}{2m}+\frac{k^{2}}{2m},
\label{3.14}
\end{eqnarray}
where the angular frequency of the harmonic oscillator is now given by $\omega=\omega_{n,\,l,\,s}$ due to the condition $a_{n+1}=0$. The condition $a_{n+1}=0$ allows us to obtain a expression involving the angular frequency $\omega$ and the quantum numbers $n$, $l$ and $s$ \cite{eug}. By writing $\omega=\omega_{n,\,l,\,s}$, we have that the choice of the values of $\omega$ depends on the quantum numbers $n$, $l$ and $s$ in order to satisfy the condition $a_{n+1}=0$. Thereby, we are assuming that $\omega=\omega_{n,\,l,\,s}$ can be adjusted in order that the condition $a_{n+1}=0$ can be satisfied. In the following, let us calculate the values of the angular frequency for $n=1$ and $n=2$. For $n=1$, we have
\begin{eqnarray}
\omega_{1,\,l,\,s}=\frac{\delta^{2}}{2m\bar{\alpha}}=\frac{\left[2gbB_{0}\nu_{s}+sgbB_{0}\right]^{2}}{2m\left(2\left|\nu_{s}\right|+1\right)},
\label{3.15}
\end{eqnarray}
and for $n=2$
\begin{eqnarray}
\omega_{2,\,l,\,s}=\frac{\delta^{2}}{4m\left(2\bar{\alpha}+1\right)}=\frac{\left[2gbB_{0}\nu_{s}+sgbB_{0}\right]^{2}}{4m\left(4\left|\nu_{s}\right|+3\right)}.
\label{3.16}
\end{eqnarray}

Hence, we have obtained the energy levels for bound states solutions of the Schr\"odinger-Pauli equation (\ref{3.1}) for the harmonic oscillator under the influence of a Coulomb-like potential induced by the Lorentz symmetry violation background. We have seen in the previous section that bound states solutions can be achieved only if the Coulomb-like potential is attractive. In contrast, we can see in (\ref{3.14}) that the energy levels of bound states can be obtained for both signs of the parameter $\delta$ given in (\ref{2.7}), in other words, the bound state solutions of the harmonic oscillator can be obtained for both attractive and repulsive Coulomb-like potential induced by Lorentz symmetry breaking. Moreover, we can observe in (\ref{3.14}) that the degeneracy of the energy levels of the harmonic oscillator is not broken by the effects of the violation of the Lorentz symmetry. Indeed, the principal influence of the Lorentz symmetry breaking stems from the term $g\,\left(\vec{b}\times\vec{B}\right)$ and the scenario defined in (\ref{2.1}) is the dependence of the cyclotron frequency on the quantum numbers $n$, $l$ and $s$. The dependence of the cyclotron frequency on the quantum numbers $n$, $l$ and $s$ means that not all values of the cyclotron frequency are allowed, but a discrete set of values for the cyclotron frequency \cite{eug,f1}. The other contribution of the Lorentz symmetry breaking to the energy levels of the harmonic oscillator is given by a new term $\frac{\left(gbB_{0}\right)^{2}}{2m}$. This behaviour of the bound states solutions of the harmonic oscillator is quite unusual because it is due to a vacuum contribution, i. e., a neutral particle confined to the harmonic oscillator potential in presence of this environment defined in (\ref{2.1}) feels this particular vacuum polarization.

\section{conclusions}

In this work, we have discussed the arising of bound states solutions for the Schr\"odinger-Pauli equation based on the effects of the Lorentz symmetry violation background. We have shown that the presence of a fixed space-like vector field parallel to the radial direction and a uniform magnetic field on the $z$ direction provides a new possible scenario of studying Lorentz symmetry breaking effects by inducing a Coulomb-like potential in which bound states solutions for the Schr\"odinger-Pauli equation can be achieved. We also have discussed the influence of this Lorentz symmetry violation background on the two-dimensional harmonic oscillator, and shown the dependence of the energy levels of the harmonic oscillator on the Lorentz violating terms. Moreover, we have shown that Lorentz symmetry breaking effects do not break the degeneracy of the energy levels of the harmonic oscillator, but impose a dependence of the cyclotron frequency on the quantum numbers $n$, $l$ and $s$. By comparing with a recent study of Lorentz symmetry breaking effects whose scenario is defined by a term which plays the role of a scalar potential \cite{bbs3}, the case discussed in this work brings a new point of view of studying Lorentz symmetry breaking effects given by a term that plays the role of a vector potential. However, one should note that detecting Lorentz symmetry violation effects at low energies is a problem in which any expected value given in terms of vacuum is weak. In recent years, studies have established limits of energy in which this breaking cannot be seen \cite{belich1,belich2,belich3}.

We would like to thank CNPq (Conselho Nacional de Desenvolvimento Cient\'ifico e Tecnol\'ogico - Brazil) for financial support.


\begin{thebibliography}{99}


\bibitem{extra3} V. A. Kostelecky and S. Samuel, Phys. Rev. Lett. \textbf{63}, 224 (1989); 
								V. A. Kostelecky and S. Samuel, Phys. Rev. Lett. \textbf{66}, 1811 (1991); 
								V. A. Kostelecky and S. Samuel, Phys. Lett. B \textbf{381}, 89 (1996); 
								V. A. Kostelecky and R. Potting, Phys. Rev. D \textbf{51}, 3923 (1995).

\bibitem{extra1} D. Mattingly, Living Rev. Relativ. \textbf{8}, 5 (2005).

\bibitem{extra2} V. A. Kostelecky, \textit{CPT and Lorentz Symmetry} (World Scientific, Singapore, 2011).

\bibitem{Colladay} D. Colladay and V. A. Kostelecky, Phys. Rev. D \textbf{55}, 6760 (1997); 
									D. Colladay and V. A. Kostelecky, Phys. Rev. D \textbf{58}, 116002 (1998); 
									S. R. Coleman and S. L. Glashow, Phys. Rev. D \textbf{59}, 116008 (1999).

\bibitem{Jackiw} S. M. Carroll, G. B. Field and R. Jackiw, Phys. Rev. D \textbf{41}, 1231 (1990).

\bibitem{KM3} V. A. Kostelecky and M. Mewes, Phys. Rev. Lett. \textbf{97}, 140401 (2006).

\bibitem{Kostelec} V. A. Kostelecky and M. Mewes, Phys. Rev. D \textbf{80}, 015020 (2009).

\bibitem{Susy} M. S. Berger and V. A. Kostelecky, Phys. Rev. D {\bf65}, 091701 (2002);
							H. Belich , J. L. Boldo, L. P. Colatto, J. A. Helay\"el-Neto and A. L. M. A. Nogueira, Phys. Rev. D \textbf{68}, 065030 (2003); 
							A. P. Baeta Scarpelli, H. Belich, J. L. Boldo, L. P. Colatto, J. A. Helay\"el-Neto and A. L. M. A. Nogueira, Nucl. Phys. Proc. Suppl.\textbf{127}, 105 (2004).

\bibitem{Cherenkov2} B. Altschul, Nucl. Phys. B \textbf{796}, 262 (2008);
										C. Kaufhold and F. R. Klinkhamer, Phys. Rev. D \textbf{76}, 025024 (2007).

\bibitem{Radiative} R. Jackiw and V. A. Kostelecky, Phys. Rev. Lett. \textbf{82}, 3572 (1999); 
										J. M. Chung and B. K. Chung Phys. Rev. D \textbf{63}, 105015 (2001); 
										G. Bonneau, Nucl.Phys. B \textbf{593}, 398 (2001); 
										M. Perez-Victoria, Phys. Rev. Lett. \textbf{83}, 2518 (1999); 
										A. P. B. Scarpelli, M. Sampaio, M. C. Nemes, and B. Hiller, Phys. Rev. D \textbf{64}, 046013 (2001); 
										A. P. B. Scarpelli, M. Sampaio, M. C. Nemes, B. Hiller, Eur. Phys. J. C \textbf{56}, 571 (2008); 

\bibitem{Casimir} M. Frank and I. Turan, Phys. Rev. D {\bf74}, 033016 (2006); 
									O. G. Kharlanov, V. C. Zhukovsky, Phys. Rev. D {\bf81}, 025015 (2010).

\bibitem{CMBR} J.-Q. Xia, H. Li, X. Wang and X. Zhang, Astron. Astrophys. \textbf{483}, 715 (2008); 
								P. Cabella, P. Natoli, J. Silk, Phys. Rev. D \textbf{76}, 123014 (2007).

\bibitem{Petrov} M. Gomes, T. Mariz, J. R. Nascimento, A. Yu. Petrov, A. F. Santos and A. J. da Silva, Phys. Rev. D \textbf{81}, 045013 (2010).

\bibitem{Adam} C. Adam and F. R. Klinkhamer, Nucl. Phys. B \textbf{607}, 247 (2001); 
								A. P. Baeta Scarpelli, H. Belich, J. L. Boldo and J. A. Helay\"el-Neto, Phys. Rev. D \textbf{67}, 085021 (2003).

\bibitem{Soldati} A. A. Andrianov and R. Soldati, Phys. Rev. D \textbf{51}, 5961 (1995); 
									J. Alfaro, A. A. Andrianov, M. Cambiaso, P. Giacconi and R. Soldati, Int. J. Mod. Phys. A \textbf{25}, 3271 (2010); 
									V. C. Zhukovsky, A. E. Lobanov and E. M. Murchikova, Phys. Rev. D \textbf{73} 065016, (2006).

\bibitem{Prop2} R. Casana, M. M. Ferreira Jr., A. R. Gomes, F. E. P. dos Santos, Phys. Rev. D \textbf{82}, 125006 (2010).

\bibitem{Klinkmicro} F. R. Klinkhamer and M. Schreck, Nucl. Phys. B \textbf{848}, 90 (2011).


\bibitem{Klink3} F. R. Klinkhamer and M. Schreck, Phys. Rev. D \textbf{78}, 085026 (2008).

\bibitem{Interact2} C. D. Carone, M. Sher, and M. Vanderhaeghen, Phys. Rev. D \textbf{74}, 077901 (2006); 
 										B. Altschul, Phys. Rev. D \textbf{79}, 016004 (2009).

\bibitem{Interact3} M. A. Hohensee, R. Lehnert, D. F. Phillips and R. L. Walsworth, Phys. Rev. D \textbf{80}, 036010 (2009); 
										B. Altschul, Phys. Rev. D \textbf{80}, 091901(R) (2009).



\bibitem{tet} V. A. Kostelecky, Phys. Rev. D \textbf{69}, 105009 (2004).

\bibitem{bbs2} K. Bakke, H. Belich and E. O. Silva, Ann. Phys. (Berlin) {\bf523}, 910 (2011).

\bibitem{bb} K. Bakke and H. Belich, J. Phys. G: Nucl. Part. Phys. {\bf39}, 085001 (2012).

\bibitem{Hamilton} V. A. Kostelecky and C. D. Lane, J. Math. Phys. \textbf{40}, 6245 (1999); 
									R. Lehnert, J. Math. Phys. \textbf{45}, 3399 (2004).

\bibitem{Manojr} M. M. Ferreira Jr. and F. M. O. Moucherek, Int. J. Mod. Phys. A \textbf{21}, 6211 (2006); 
								S. Chen, B. Wang, and R. Su, Classical Quant. Grav. \textbf{23}, 7581 (2006); 
								O. G. Kharlanov and V. Ch. Zhukovsky, J. Math. Phys. \textbf{48}, 092302 (2007). 

\bibitem{Nonmini} H. Belich, T. Costa-Soares, M. M. Ferreira Jr., J. A. Helay\"el-Neto and F. M. O. Moucherek, Phys. Rev. D \textbf{74}, 065009 (2006).

\bibitem{novo} M. Frank, I. Turan and I. Yurdusen, JHEP01(2008)039

\bibitem{novo1} G. Gazzola, H. G. Fargnoli, A. P. Ba\^eta Scarpelli, M. Sampaio and M. C. Nemes, J. Phys. G: Nucl. Part. Phys. \textbf{39}, 035002 (2012).


\bibitem{ab} Y. Aharonov and D. Bohm,  Phys. Rev. {\bf115}, 485 (1959).

\bibitem{ahan} Y. Aharonov and J. Anandan, Phys. Rev. Lett. {\bf 58}, 1593 (1987).

\bibitem{berry} M. V. Berry, Proc. R. Soc. Lond. A {\bf392}, 45 (1984).

\bibitem{anan2} J. Anandan, Phys. Lett. A {\bf138}, 347 (1989); 
								J. Anandan, Phys. Rev. Lett. {\bf85}, 1354 (2000).


\bibitem{belich1} H. Belich, T. Costa-Soares, M. M. Ferreira Jr. and J. A. Helay\"el-Neto, Eur. Phys. J. C {\bf41}, 421 (2005).

\bibitem{belich} H. Belich, T. Costa-Soares, M. M. Ferreira Jr., J. A. Helay\"el-Neto, M. T. D. Orlando, Phys. Lett. B \textbf{639}, 675 (2006).

\bibitem{belich2} H. Belich, T. Costa-Soares, M. M. Ferreira Jr., J. A. Helay\"el-Neto and F. M. O. Moucherek, Phys. Rev. D \textbf{74}, 065009 (2006).

\bibitem{belich3} H. Belich, L. P. Collato, T. Costa-Soares, J. A. Helay\"el-Neto and M. T. D. Orlando, Eur. Phys. J. C \textbf{62}, 425 (2009).

\bibitem{ac} Y. Aharonov and A. Casher, Phys. Rev. Lett. {\bf 53}, 319 (1984).

\bibitem{proc} J. S. Diaz, arXiv:1109.4620 [hep-ph]

\bibitem{bbs} K. Bakke, H. Belich and E. O. Silva, J. Math. Phys. {\bf52}, 063505 (2011).

\bibitem{mano} H. Belich, E. O. Silva,  M. M. Ferreira Jr. and M. T. D. Orlando, Phys. Rev. D. {\bf83}, 125025 (2011). 

\bibitem{hmw} X. G. He and B. H. J. McKellar, Phys. Rev. A {\bf47}, 3424 (1993); 
							M. Wilkens, Phys. Rev. Lett. {\bf72}, 5 (1994).


\bibitem{greiner} W. Greiner, \textit{Relativistic Quantum Mechanics: Wave Equations, 3rd Edition} (Springer, Berlin, 2000).

\bibitem{bbs3} K. Bakke, E. O. Silva and H. Belich, J. Phys. G: Nucl. Part. Phys. {\bf39}, 055004 (2012).

\bibitem{schu} P. Schl\"uter, K.-H. Wietschorke and W. Greiner, J. Phys. A {\bf16}, 1999 (1983).

\bibitem{b4} K. Bakke, Ann. Phys. (Berlin) {\bf523}, 762 (2011); 
							K. Bakke, Int. J. Mod. Phys. A, {\bf26}, 4239 (2011);
							K. Bakke, Eur. Phys. J. Plus {\bf127}, 82 (2012).

\bibitem{bd}  N. D. Birrell and P. C. W. Davies, \textit{Quantum Fields in Curved Space}, (Cambridge University Press, Cambridge, UK, 1982).

\bibitem{weinberg} S. Weinberg, {\it Gravitation and Cosmology: Principles and Applications of the General Theory of Relativity} (IE-Wiley, New York, 1972).

\bibitem{mello} E. R. Bezerra de Mello, J. High Energy Phys. JHEP06(2004)016 


\bibitem{abra} M. Abramowitz and I. A. Stegum, \textit{Handbook of mathematical functions} (Dover Publications Inc., New York, 1965).

\bibitem{landau} L. D. Landau and E. M. Lifshitz, \textit{Quantum Mechanics, the nonrelativist theory, 3rd Ed.} (Pergamon, Oxford, 1977).

\bibitem{heun} A. Ronveaux, \textit{Heun's differential equations} (Oxford University Press, Oxford, 1995).



\bibitem{eug} E. R. Figueiredo Medeiros and E. R. Bezerra de Mello, Eur. Phys. J. C {\bf72}, 2051 (2012).

\bibitem{b6} K. Bakke, Braz. J. Phys. {\bf41}, 167 (2011); 
						K. Bakke, Ann. Phys. (Berlin) {\bf524}, 338 (2012).



\bibitem{arf} G. B. Arfken and H. J. Weber, {\it Mathematical Methods for Physicists, sixth edition} (Elsevier Academic Press, New York, 2005)

\bibitem{f1} C. Furtado, B. G. C. da Cunha, F. Moraes, E. R. Bezerra de Mello and V. B. Bezerra, Phys. Lett. A {\bf195}, 90 (1994).



\end{thebibliography}
\end{document}